\documentclass[doublecol]{epl2}

\usepackage{graphics} 
\usepackage{epsfig} 
\usepackage{amsmath} 
\usepackage{amssymb}  
\usepackage{amsfonts, color}
\usepackage{graphicx, subfigure}
\usepackage[normalem]{ulem}
\usepackage{epstopdf}

\newcommand{\vX}{\boldsymbol{X}}
\newcommand{\vx}{\boldsymbol{x}}
\newcommand{\vnu}{\boldsymbol{\nu}}
\newcommand{\vp}{\boldsymbol{p}}

\def\b{\textcolor{blue}}

\title{ {Rare Event} Extinction on {Stochastic} Networks}
\author{Brandon S.  Lindley\inst{1}, Leah B. Shaw\inst{2}, and Ira B. Schwartz\inst{1}}
\shortauthor{Lindley, Shaw, and Schwartz}
\institute{
\inst{1} US Naval Research Laboratory, Code 6792, Nonlinear System Dynamics Section, Plasma Physics Division, Washington, DC 20375\\
\inst{2} College of William and Mary, Department of Applied
  Science, Williamsurg, VA 23187-8795
}

\abstract{
 We consider the problem of  extinction processes on random networks with
  a given structure. For sufficiently large well-mixed populations, the
  process of extinction of one or more state variable  components occurs in
  the tail of the quasi-stationary probability distribution, thereby making it a rare event. Here we show
  how to  extend the theory of large deviations to random networks to predict extinction times. In particular, we use the theory to find the most probable path leading
to extinction. We apply the methodology to epidemic models and discover how mean extinction
times scale with epidemiological and network parameters in Erd\H{o}s-R\'{e}nyi networks. The
results are shown to compare quite well with Monte Carlo simulations of the
network in predicting both the most probable paths to extinction and mean extinction times.
}

\pacs{05.50.-a}{Fluctuation phenomena, random processes, noise, and Brownian motion}
\pacs{87.23.Cc}{Population dynamics}

\begin{document}
\maketitle

In many models of finite populations,
  random fluctuations occur due to internal interactions between individuals
  or agents, and/or external stochastic forces. Such fluctuations are evident in
  the modeling of well mixed populations that support epidemics of disease
  spread \cite{Anderson91}, as well as ecological bio-diversity of species
  \cite{BanMar09,Azaele2006}, among others.  Typically, the
dynamics performs small random fluctuations about an attracting state. However, in almost all
populations of finite size, fluctuations may organize in such a way to drive
one or more components of the population to extinction. Mechanisms conjectured to
play a role in extinction processes include small population size, low contact
frequency for frequency-dependent transmission, competition for
resources, and evolutionary pressure~\cite{DeCastro2005}, as well as heterogeneity in
populations and transmission in coupled population models~\cite{Lloyd2007,Eriksson2013}.

In epidemic models where the population is well-mixed, extinction of
  infectious individuals has been shown to be affected by  noise
  intensity \cite{Melbourne2008}, peak infectious population size
  \cite{Alonso2006}, and seasonal phase occurrence \cite{stolhu07}. Moreover,
  since the extinct state is typically unstable in the deterministic mean field and is an
  absorbing state of a stochastic process, time scales for extinction may be
  exponentially long \cite{dyscla08}; i.e., the probability of extinction
is a decreasing exponential function \cite{Kubo63}. Vaccination and treatment
  programs have been studied to speed up the extinction of disease in
  well-mixed populations \cite{Nasell1999, dyscla08}. For example, although  most
  vaccination schedules are designed to be administered periodically
  (deterministic) \cite{bolker1996impact,schwartz2004,Dykman_vacc},
  Poisson distributed scheduling was recently shown to be more efficient than regular
  treatment schedules \cite{billings2013intervention}.

In network populations, outbreak extinction probabilities have been predicted for early times when an
infection has just been introduced \cite{Keeling2005,Trapman2007}.  Other
studies of extinction on networks attempt to predict whether a persistent
non-extinct state exists, such as for computer viruses in growing networks
\cite{Hayashi2004} and epidemics in various network geometries (e.g.,
\cite{Silva2007}).  This is equivalent to finding the bifurcation point where
the extinct state becomes unstable and thus can frequently be predicted using
deterministic approximations (e.g., \cite{Hayashi2004}).

 Extinction times for
an endemic disease in a network have occasionally been studied, mostly numerically. In
\cite{Shaw2008a}, extinction times were measured in simulations of an epidemic
on an adaptive network, and the log of the extinction time was observed to
increase with a power of the distance from a bifurcation point.  In
\cite{Shaw2010}, lifetimes were measured in a similar model with the addition
of pulsed vaccination.  Similar lifetimes were obtained in simulations by
adding stochastic pulsed vaccination to a deterministic mean-field model based
on a pair approximation, suggesting that it is not necessary to model the full
stochastic network dynamics in detail to capture the extinction time.  Here we
will present a method for finding the most probable path to extinction and the
extinction time in networks that can be described using a pair
approximation.

Here we consider the problem of epidemic extinction in stochastic networks,
 and we find that the extinction process depends not just on the nodes of
  the network, but also on how the  links change as the system evolves. That
  is, along the most probable path to extinction, we have derived a new approximate model showing that both nodes and links play an
important role in  the mean time to disease eradication. We have introduced a novel mathematical tool so that
the path is derived constructively.


The specific example we will consider here is a network of $N$
nodes and $K$ links, with an average degree of $2K/N$. The dynamics on the
network is  an \textbf{SIS} epidemic model, where susceptibles capable of
acquiring the disease become infectious through a contact with an infective
individual, and become susceptible again after a recovery period.
Here we divide the population into two groups,  with $S$ as the number of susceptible
individuals and $I$ as the number of infected individuals, such that
$S+I=N$. The population is closed, and there are no births and
  deaths.

To quantify how the network topology compares with a well-mixed global, or
  all-to-all, coupling structure, we allow for two modes of disease spread. Specifically, global disease
transmission occurs in a well-mixed population when the disease transmits from any infected person to any
susceptible person. In contrast to global transmission models, local disease
transmission occurs when a network of
connections between individual persons is considered, and diseases can only
spread along links between infected and susceptible individuals.
In the case of transmission via only the network, we let  $p$  be the infection rate per \textbf{SI} link.   When transmission is only global, we define a global contact rate of $p2K/N$, selected so that the epidemic threshold occurs at $p^*=\frac{rN}{2K}$ in both the globally coupled and network transmission cases.    We  will introduce a homotopy parameter $\epsilon$ which continuously
transforms the system between
the local and  global transmission models.  Finally, $r$ is a recovery rate.

We approximate the dynamics of the network by considering transitions in nodes and links, similar to the pair-based proxy model of \cite{Rogers2012}. We let $\vX$ denote the vector with components containing both
  the number of nodes and number of links of the network. Specifically,
    we let $\vX=[S,I,N_{SS},N_{SI},N_{II}]$, where $S,I$ denotes the numbers of
    susceptibles and infectives, respectively, and  $N_{AB}$ denotes the number of connections
between  nodes of type \textbf{A} and \textbf{B}.  We assume large
  but finite population size, $N$, and we suppose the dynamics proceeds as a
  Markov process. To complete the formulation for the network dynamics, we
  suppose there exist $M$ transition events with transition rates
  $W(\vX,\vnu_k)$ having increments $\vnu_k$. The
  dynamics of the probability density, $\rho(\vX,t)$, can now be modeled as a master equation \cite{Kubo63,vanKampen_book,gang87}:
\begin{equation}
\begin{split}
\frac{\partial\rho}{\partial t}(\vX,t) &= \sum_{k=1}^{M}[W(\vX-\vnu_k,\vnu_k)\rho(\vX-\vnu_k,t)\\
      &\quad -W(\vX,\vnu_k)\rho(\vX,t)], \label{mastereq}
\end{split}
\end{equation}
where $\rho(\vX,t)$ represents the probability of finding the system in state
$\vX$ at time $t$.

The rare events are characterized by observing the targeted event in the
  tail of the probability distribution. Typically, when observing the times for the event
  to occur, one sees that the distribution of times is exponential \cite{sbdl09}. When the
  population is sufficiently large, we may assume the distribution of
  times possesses such an exponential tail \cite{dyscla08}.

Making a homogeneity assumption that each node's local neighborhood is identical to that of all other nodes of the same type, we can find its neighbors from the system average. For example, a node of type \textbf{S}
would have an expected degree $d_S=(2N_{SS}+N_{SI})/S$ and has, on average,
$2N_{SS}/S$ neighbors of type \textbf{S}.

We next specify the possible transitions and associated transition rates for the pair-based proxy model, so that the approximate stochastic process can be described by a master equation that captures fluctuations in the probability density.
There are three types of state transitions to consider: \textbf{S}$\rightarrow$\textbf{I}
by transmission along the network, \textbf{S}$\rightarrow$\textbf{I} by global disease
transmission, and \textbf{I}$\rightarrow$\textbf{S} by disease recovery.

To determine the increments for each  transition, notice that whenever a node changes from one state to another, all of its associated links
change. For example, if a susceptible node becomes infected via local disease
transmission, then \textbf{S}$\rightarrow$\textbf{I}, and therefore all its \textbf{SS} links become
\textbf{SI} links.  Further, $1+N_{SI}/S$ links of
type \textbf{SI} will be lost, creating as many \textbf{II} links.  Using this logic, we can
write the following transition increments for transmission along the network, global transmission, and recovery, respectively: 
\begin{equation}
\label{eq:transitions}
\begin{aligned}
\vnu_1 &=\left[-1,1,-\frac{2N_{SS}}{S},\frac{2N_{SS}}{S}-\left(1+\frac{N_{SI}}{S}\right),\left(1+\frac{N_{SI}}{S}\right)\right] &&\\
\vnu_2 &=\left[-1,1,-\frac{2N_{SS}}{S},\frac{2N_{SS}}{S}-\frac{N_{SI}}{S},\frac{N_{SI}}{S}\right] &&\\
\vnu_3 &=\left[1,-1,\frac{2N_{SI}}{I},-\frac{N_{SI}}{I}+\frac{2N_{II}}{I},-\frac{2N_{II}}{I}\right]. &&
\end{aligned}
\end{equation}

For each transition $\vnu_k$ we have an associated transition rate $W(\vX,\vnu_k)$ given by
\begin{equation}
\label{eq:transitionrates}
\begin{aligned}
W(\vX,\vnu_1) &=\epsilon p N_{SI}&&\\
W(\vX,\vnu_2) &=(1-\epsilon)p \frac{2K}{N}\frac{\b{S I}}{N} &&\\
W(\vX,\vnu_3) &=r\b{I}, &&
\end{aligned}
\end{equation}
for transmission along the network, global transmission, and recovery, respectively.  We have
introduced the homotopy parameter $\epsilon$ which continuously
transforms the system between
the  global and network transmission models.  Here $\epsilon=0$ returns the
standard equations for the  \textbf{SIS}  model in a globally-coupled (well
mixed) population, and $\epsilon=1$ is a locally-coupled network \textbf{SIS} model. Even the pair-based proxy system is sufficiently high dimensional that finding the most probable path to extinction will be difficult a priori, but continuously varying $\epsilon$ will allow us to track the path from the known case of global coupling to the network case.  This approach to tracking the path to extinction by continuously varying a parameter is new. However, we 
note that far from the extinction bifurcation point, if the well mixed population is
constrained in fluctuation so that $S+I=N$, the limit of $I(\epsilon)$ as $\epsilon$ approaches
zero may not exist since the \textbf{SIS} model has been shown to be fragile \cite{Khasin2009}.

The rare events we are interested in  are those of extinction where the
  number of {infected} nodes goes to zero. As in the globally coupled case ($\epsilon=0$),
  the Monte Carlo dynamics of the network ($\epsilon=1$) exhibits random fluctuations about
  an attracting state, and then the fluctuations organize in such a way as to
  drive the infected {nodes} to extinction. We wish to characterize {the}
  probability of an extinction event in the large population limit and
  compare it to the globally coupled case. In particular, we are interested in the
  most probable path and mean times to extinction. To understand the scaling of extinction
  times in terms of epidemiological parameters and network topology, we employ
large deviation theory techniques for finite populations \cite{Dykman,gang87}.


We introduce the scaled variables $\vx$ with components defined by
$x_1=I/N$, $x_2=N_{SI}/N$, and $x_3=N_{II}/N$, where the remaining
 variables $S$ and $N_{SS}$ can be eliminated by conservation of nodes and links. We define the scaled
transition rates $w(\vx,\vnu_k)=W(\vX,\vnu_k)/N$ on these
variables. Henceforth, we will refer to the fraction of infecteds and
susceptibles as $P_I=I/N$ and $P_S=S/N$. The link {variables} will be given as
{$L_{AB}=N_{AB}/N$}. Given that the probability current at the extinct state is
sufficiently small {for large enough $N$,} and that the increments are sufficiently small with
respect to $N$, there will exist a quasi-stationary probability distribution
about a state with a non-zero number of infected individuals that decays into
the stationary (extinct) solution over exponentially long times
\cite{billings2013intervention}. The extinction rate of infected individuals can be calculated from the
tail of the quasi-stationary distribution.  It has been demonstrated that a WKB
approximation to the quasi-stationary distribution allows one to approximate
the mean extinction time provided the population is sufficiently large when $\epsilon=0$
\cite{Dykman}.

Employing the WKB approximation for the probability, $\rho(\vx,t)=A(\vx)\text{exp}(-NR(\vx,t))$, from the first order expansion with respect to $N^{-1}$, we can write the Hamilton-Jacobi equation,
\begin{equation}
\label{Ham-Jac}
\frac{\partial R}{\partial t}+H\left(\vx,\frac{\partial R}{\partial \vx}\right)=0.
\end{equation}
Here the function $H$ is called the Hamiltonian of the system, and {$R$} is known as the action. The variable $\vp=\partial_x R$ is the conjugate momenta of the Hamilton-Jacobi equation, and the Hamiltonian can be written  as
\begin{equation}
H(\vx,\vp) = \sum_{k=1}^{3}w(\vx,\vnu_k)(e^{\vp\cdot\vnu_k}-1).
\end{equation}
  We  analyze the system by solving the characteristic equations
$\dot{\vx}=\partial_{\vp}H(\vx,\vp),$ $\dot{\vp}=-\partial_{\vx}
H(\vx,\vp)$.

Since we are  interested in the probability of extinction events, we must
compute the action $R$ that maximizes this probability. This will, naturally,
be the one that satisfies the Hamilton-Jacobi Eq.~\ref{Ham-Jac}. This
implies that there will be some trajectory along which $R$ is minimized, which
represents the maximal probability of such an extinction event, and that this
trajectory, which we will call the ``most probable path,'' satisfies the
characteristic equations for the Hamiltonian plus boundary conditions \cite{Dykman,gang87}.

The boundary conditions for the characteristic equations are given as steady
{state} solutions of $\dot{\vx}=\dot{\vp}=0$. In fact, for epidemic models, the
quasi-stationary probability distribution peaks at an endemic state, where the
number of new infections equals the number of recoveries per unit time, and
corresponds to the zero conjugate momenta case, i.e.,
the steady state
  is at $(\vx,\vp)=(\vx_e,\bold{0}$). The other steady state is the extinct state, $P_S=1,$ $P_I=0$, {which is saddle point.}  However, here we find that
the conjugate momenta at this state, $\vp_0$, is non-zero, since there is a non-zero
  probability current at that point.  We
 label the endemic state as $(\vx_e,0)$, and the non-trivial extinct state as
$(\vx_0,\vp_0)$. The steady states of the characteristic
equation corresponding to zero conjugate momenta are those which satisfy the
deterministic epidemic model.

For the case $\epsilon=0$, analytic solutions are available for the 
most probable path to extinction \cite{Meerson2008}, and hence the action may be computed directly. However,
for $\epsilon=1$, the additional equations for the mean link fractions
complicate the analysis, and the  most probable path must be computed numerically.
We  therefore employ the iterative action minimizing method (IAMM) \cite{lindley2013iterative} to compute the
most probable path between $\vx_e$ and $\vx_0$, and the action will be given by the path
integral $\int_{\vx_e}^{\vx_0}\vp \cdot d\vx$ along this path since $H(\vx,\vp)=0$ along the path.

\begin{figure}[h]
\begin{minipage}{0.49\linewidth}
\includegraphics[scale=.18]{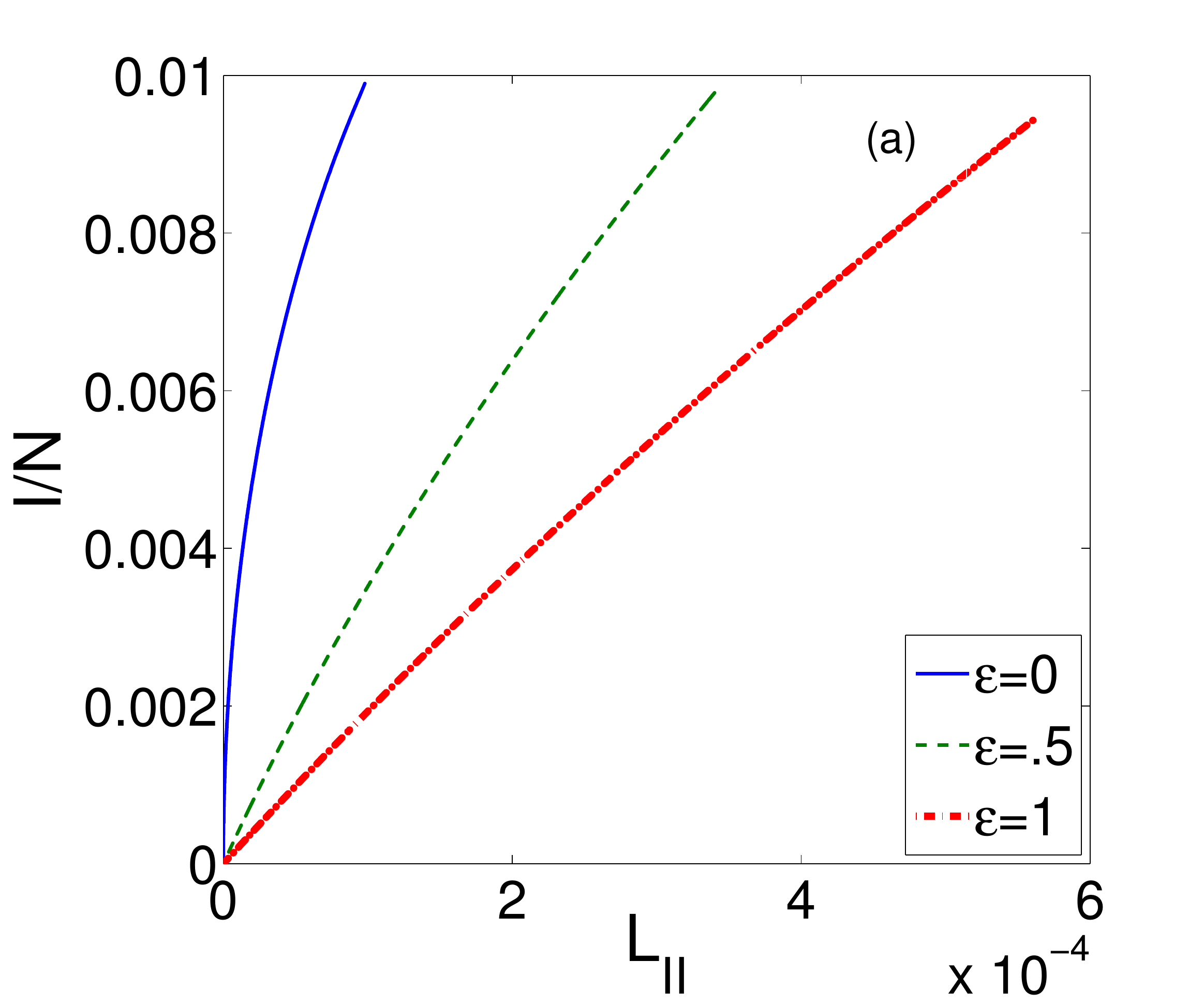}
\end{minipage}
\begin{minipage}{0.49\linewidth}
\includegraphics[scale=.18]{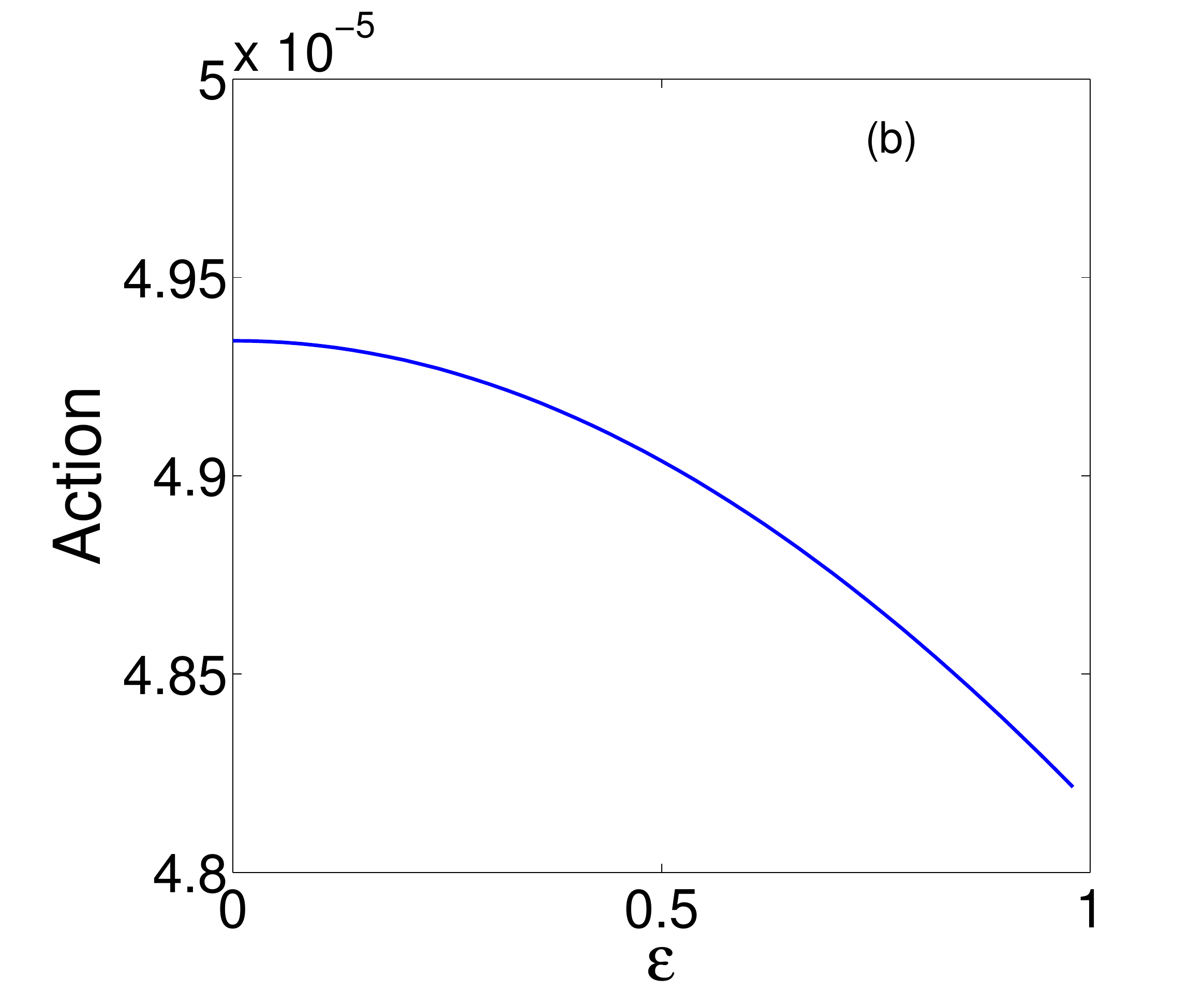}
\end{minipage}
\caption{(a) The  most probable path  projected onto the $I/N$
  versus $L_{II}$ axis for various $\epsilon$ values up to $\epsilon=1$. (b)
  The action predicted by the IAMM along the most probable path as a function of the
  continuation parameter $\epsilon$. Parameters:
  $p=1.03\times10^{-4}$, $r=0.002$ and  $K/N=10$. }\label{OP_epsilon}
\end{figure}

Using the IAMM, we perturb off the $\epsilon=0$ solution,  where the path
  can be analytically defined.  For an all-to-all connected graph,
  $L_{SI}=\frac{K}{N} 2 P_S P_I$ and $L_{II}=\frac{K}{N}P_I^2$ and momenta
  conjugate to the link variables are 0. We then use continuation as a
  function of $\epsilon$ to get to  the locally-coupled $\epsilon=1$ case.  We
  demonstrate the robustness of this approach in Fig.~\ref{OP_epsilon},
  where the computed most probable path is plotted for several
  values of $\epsilon$. It is clear from Fig.~\ref{OP_epsilon}a that the
  path predicted for the discretely coupled network ($\epsilon=1$) is quite
  different from the globally coupled disease dynamics ($\epsilon=0$). Note
    that as $\epsilon$ increases and infection spreads primarily along the
    network, \textbf{II} connections become more prominent because infected nodes
    arise next to other infected nodes.  Also, as shown in Fig.~\ref{OP_epsilon}b, the action decreases, corresponding to a decrease in
    the extinction time, presumably due to stochastic effects in local
    neighborhoods of nodes.

\begin{figure}[h]
\begin{minipage}{0.48\linewidth}
\hspace{-0.9cm}
\includegraphics[scale=.16]{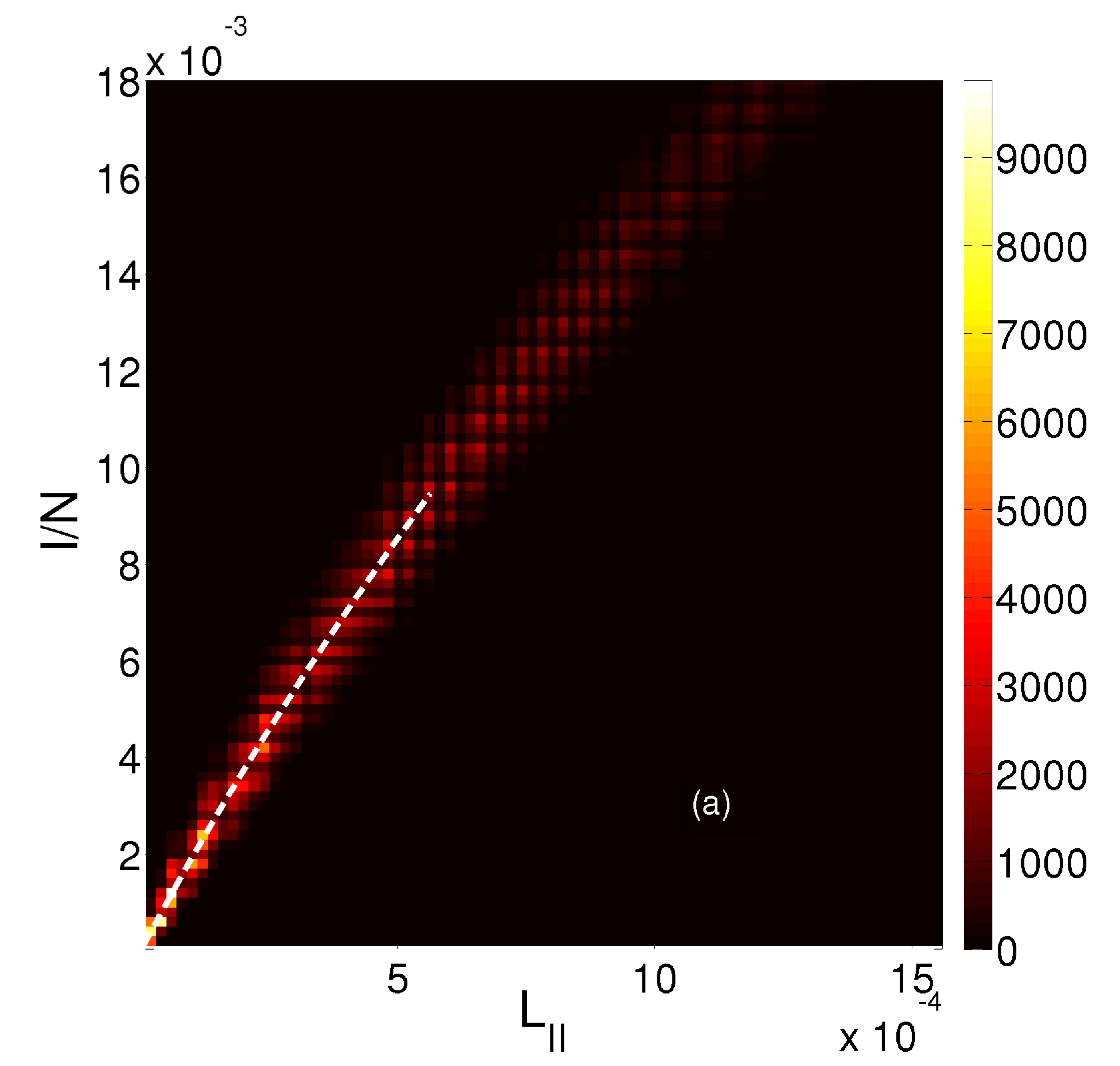}
\end{minipage}
\begin{minipage}{0.48\linewidth}
\begin{center}
\includegraphics[scale=.16]{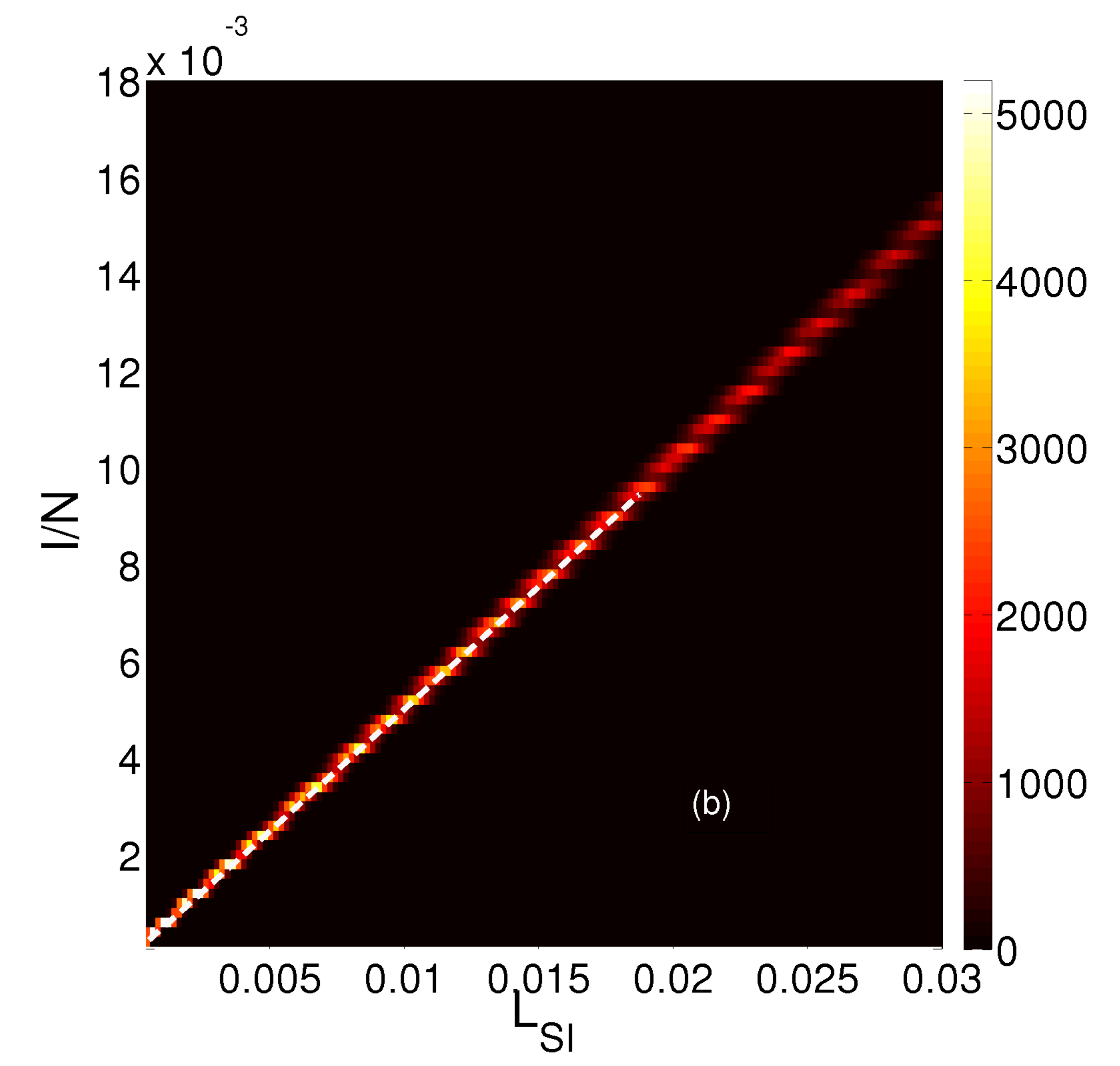}
\end{center}
\end{minipage}
\caption{(a) The position density function of extinction paths computed from stochastic simulations on Erd{\H o}s-R\'{e}nyi random networks, projected onto the $I/N$ versus $L_{II}$ axis, and overlaid with the predicted  most probable path (dashed curve).  (b) The same graph as (a) but projected onto $I/N$ versus $L_{SI}$. Parameters: $p=1.03\times10^{-4}$, $r=0.002$, $N=10^4$, $K=10^5$, and $\epsilon=1$. }\label{MCvsIAMM}
\end{figure}

To validate the predicted paths and extinction times, we compare the
numerical results for our approximate system to Monte Carlo simulations of  an
 \textbf{SIS} epidemic model on an
Erd\H{o}s-R\'{e}nyi random network. Here, the Monte Carlo simulations are
completed using
the Gillespie algorithm \cite{Gillespie1977} with an initial condition at the endemic steady state.  For the networks, 1000 trajectories are run to extinction, 100 for each of 10 randomly generated network geometries.  From the set of paths that go extinct, a density function is created from the prehistory of these paths,
and a clear local maximum  can be identified. This maximum corresponds to the most likely trajectory connecting
the endemic and extinct points, and  is shown in the density plots of
Fig.~\ref{MCvsIAMM}. Using the IAMM to compute the  most probable path and comparing it to the
 prehistory of extinction events on stochastic networks, as shown by the dashed curve in
  Fig.~\ref{MCvsIAMM}, demonstrates that our proxy model approximates well  the path to extinction.

From Monte-Carlo simulations, we can also approximate the mean lifetime of the
disease from endemic state until extinction.  We assume that the mean lifetime $\tau$ is inversely
proportional to the probability of the event, and thus can be related to our
action by the formula $\tau=B(\vx)e^{NR}$. Note the appearance of the
 prefactor $B(\vx)$ in this calculation. The prefactor may
depend on all parameters for a given problem, but in general scales as $\frac{1}{\sqrt{N}}$ for sufficiently large $N$. For our purposes, the prefactor can be found analytically for $\epsilon=0$,
\begin{equation}
B = \frac{\sqrt{2\pi\frac{R_{0}^{\text{eff}}}{N}}}{r(R_{0}^{\text{eff}}-1)^2}, \label{eq:prefactor}
\end{equation}
where the effective reproductive number is given as
$R_{0}^{\text{eff}}=2pK/(Nr)$ \cite{assaf2010,billings2013intervention}. 

\begin{figure}[h]
\begin{minipage}{0.49\linewidth}
\includegraphics[scale=.175]{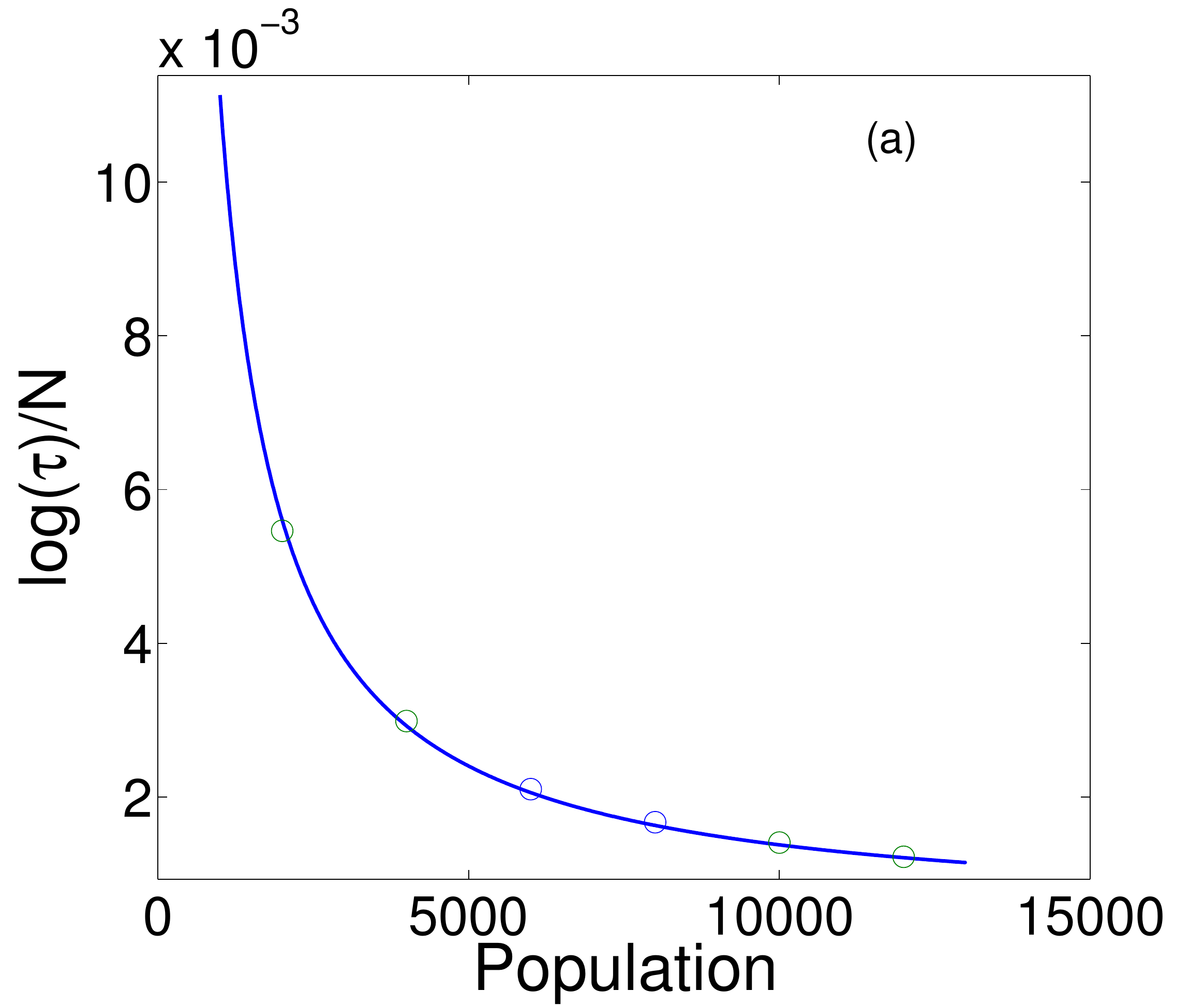}
\end{minipage}
\begin{minipage}{0.49\linewidth}
\includegraphics[scale=.175]{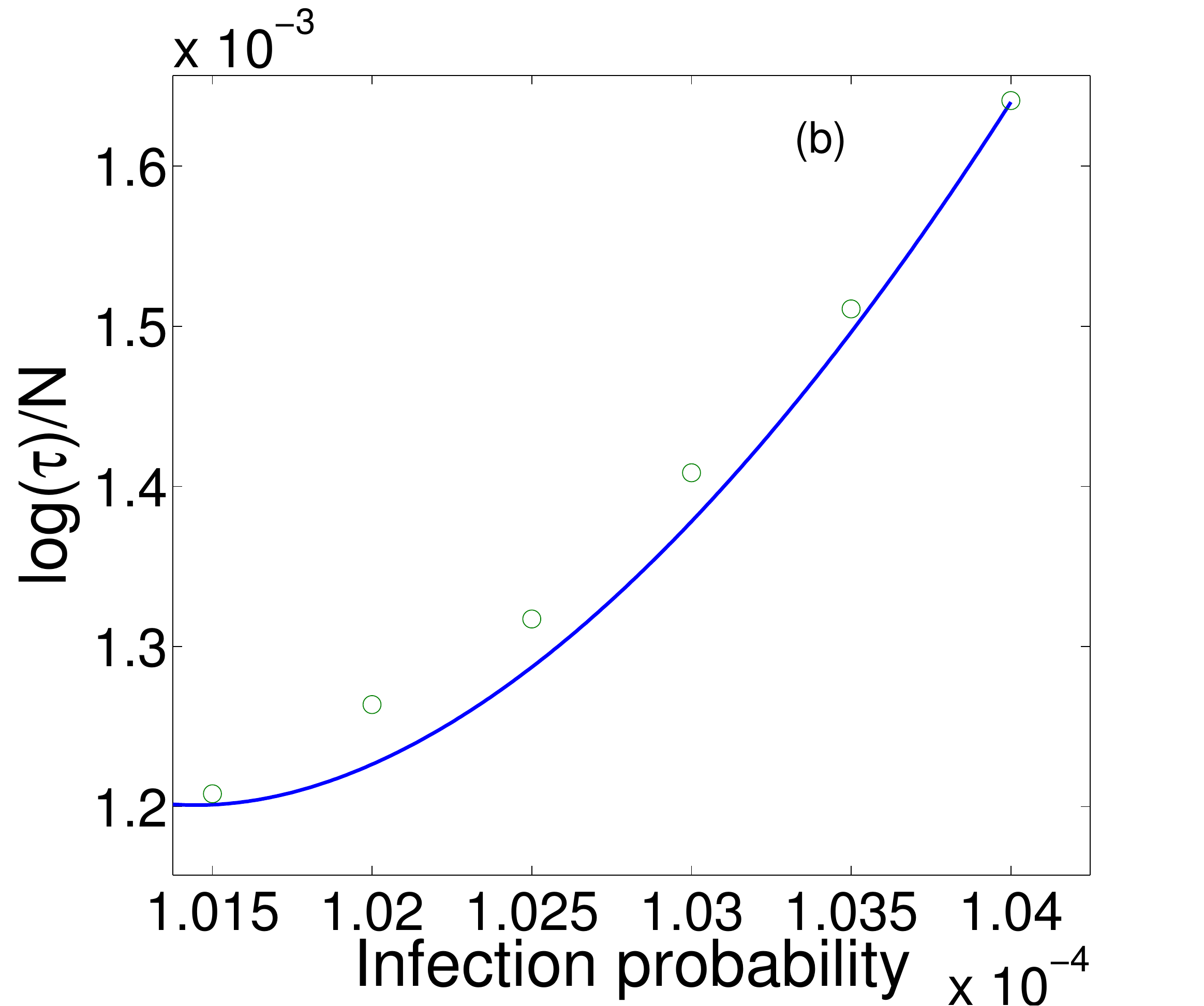}
\end{minipage}
\caption{a) $\ln(\tau)/N$ versus $N$ predictions  (solid curve) compared to mean over 1000 realizations of
  extinction on a random network  (circles) for $r=0.002$,
  $p=1.03\times10^{-4}$, $K/N=10$, and $\epsilon=1$.  b)  $\ln(\tau)/N$
  versus $p$ predictions (solid curve) compared to mean over 1000 network
  extinction events (circles) for $r=0.002$, $K=10^5$, $N=10^4$, and
  $\epsilon=1$. Note there is no fitting parameter in the theoretical plots.   }\label{ActionScaling}
\end{figure}

{In} Fig.~\ref{ActionScaling}, we plot the log of the mean lifetimes of the
disease compared to our numerical predictions of the action, incorporating the
prefactor in Eq.~\ref{eq:prefactor}. Because it is analytically intractible to find the prefactor for $\epsilon> 0$, we assume that, since the action does not vary greatly with respect to $\epsilon$, the change to the prefactor will be negligible, and thus we can use the same prefactor for the case when $\epsilon=1$.

Figure \ref{ActionScaling}a shows the log of the mean lifetime versus
population for Monte Carlo simulations of the disease on discrete networks
($\epsilon=1$) and the mean-field prediction generated by the IAMM, and scaled
by the analytical prefactor, i.e., $\ln{\tau}/N\approx R+\ln{B}/N$. Because
the action $R$ is invariant with respect to $N$, the scaling depends entirely
on the prefactor, and the good agreement between our analytical approximation
and stochastic simulations shows that our approximation of the prefactor is
sufficient to capture the dynamics. Figure \ref{ActionScaling}b varies a
different parameter, the infection rate $p$, and shows the {lifetime}
scaling predicted by our model and the stochastic simulations as the
probability of disease propagation along the network is increased.

We have presented a method for predicting extinction in stochastic network
systems by analyzing a pair-based proxy model. Both extinction times and paths
to extinction were obtained.  Tracking the path to extinction was aided by
perturbing from the known path in a well-mixed system.  In the future, the
pair-based proxy method will be extended to systems such as epidemics on
adaptive networks (e.g., \cite{Gross2006}) by adding network adaptation to the
list of transitions (Eqs.~\ref{eq:transitions} and \ref{eq:transitionrates}).
Vaccination transitions can also be added, as in \cite{Shaw2010}.  More
generally, this pair-based proxy method will be applicable to predict
extinction in any network system that is well described by a pair-based
approximation for dynamics of nodes and links, including games on networks
(e.g., \cite{Rand1999,Ohtsuki2006}).   Further, we expect that our method of
continuously varying a parameter while tracking the path to extinction will be
useful in other contexts where finding the path a priori is difficult due to
high dimensionality, {as in finite mode projection of partial differential
equations \cite{Bunin2012}, and pattern switching along paths in continuous
systems \cite{WeinanE2004}.}

BL was a National Research Council post doctoral fellow. IBS was
supported by NRL base funding (N0001414WX00023) and Office of Naval Research (N0001414WX20610). LBS  was supported by the Army Research Office, Air Force Office of Scientific
Research, and by Award Number R01GM090204 from the National Institute of General Medical Sciences.


%

\end{document}